\documentclass[preprint,trackchanges]{aastex631}

\pdfoutput=1
\usepackage{bm,amsmath}
\shorttitle{Draft}
\shortauthors{Cheng et al.}

\graphicspath{{./}{figures/}}

\begin{document}

\title{Simulation of the Solar Energetic Particle
 Event on 2020 May 29 Observed by Parker Solar Probe}

\author[0000-0002-3824-5172]{Lei Cheng}
\affiliation{Department of Aerospace, Physics and Space Sciences, Florida Institute of Technology, 150 W. University Blvd.
Melbourne, FL 32901, USA}

\author[0000-0003-3529-8743]{Ming Zhang}
\affiliation{Department of Aerospace, Physics and Space Sciences, Florida Institute of Technology, 150 W. University Blvd. 
Melbourne, FL 32901, USA}

\author[0000-0002-3176-8704]{David Lario}
\affiliation{ NASA, Goddard Space Flight Center, Heliophysics Science Division, 8800 Greenbelt Rd. Greenbelt, MD 20771, USA}

\author[0000-0003-1162-5498]{Laura A. Balmaceda}
\affiliation{ NASA, Goddard Space Flight Center, Heliophysics Science Division, 8800 Greenbelt Rd. Greenbelt, MD 20771, USA}
\affiliation{George Mason University, 4400 University Dr. Fairfax, Virginia 22030, USA}
\author[0000-0002-2106-9168]{Ryun Young Kwon}
\affiliation{Korea Astronomy and Space Science Institute, Daedeokdae-ro 776, Yuseong-gu Daejeon 34055, Republic of Korea}

\author[0000-0002-0978-8127]{Christina Cohen}
\affiliation{Space Radiation Laborotry, California Institute of Technology, Pasadena, CA 91125, USA}

\correspondingauthor{Lei Cheng}
\email{lcheng@fit.edu}

\begin{abstract}

This paper presents a stochastic three-dimensional (3D) focused transport simulation of solar energetic particles (SEPs) produced by a data-driven coronal mass ejection (CME) shock propagating through a data-driven model of coronal and heliospheric magnetic fields. The injection of SEPs at the CME shock is treated using diffusive shock acceleration of post-shock superthermal solar wind ions. A time backward stochastic simulation is employed to solve the transport equation to obtain the SEP time-intensity profile at any location, energy, and pitch angle. The model is applied to a SEP event on 2020 May 29, observed by \textit{STEREO-A} close to $\sim$1 au and by \textit{Parker Solar Probe} (\textit{PSP}) when it was about 0.33 au away from the Sun. The SEP event was associated with a very slow CME with a plane-of-sky speed of 337 $\rm km\;s^{-1}$ at a height below 6 $\rm R_S$ as reported in the SOHO/LASCO CME catalog. We compute the time profiles of particle flux at \textit{PSP} and \textit{STEREO-A} locations, and estimate both the spectral index of the proton energy spectrum for energies between $\sim$2 and 16 MeV and the equivalent path length of the magnetic field lines experienced by the first arriving SEPs. We found that the simulation results are well correlated with observations. The SEP event could be explained by the acceleration of particles by a weak CME shock in the low solar corona that is not magnetically connected to the observers.
\end{abstract}


\section{Introduction} \label{sec:intro}

Solar energetic particles (SEPs) consist of electrons, protons, and heavy ions produced in association with solar eruptions that occasionally can reach up to GeV energies. They have been studied for several decades since they can be directly measured by particle detectors on spacecraft and indirectly by neutron monitors on the ground during ground-level enhancement events. Understanding the origin and transport of SEPs is of vital importance to space weather predictions since exposure to a large number of high-energy particles could pose a significant risk to spacecraft electronics and astronauts in space.

SEPs are believed to be produced by either magnetic reconnection in solar flares or particle acceleration at shocks driven by coronal mass ejections (CMEs). Historically, SEP events have been divided into impulsive SEP events associated with solar flares, and gradual SEP events whose particles are thought to be accelerated by CME-driven shocks. Gradual events have proton intensities that are usually more elevated and longer-lived that those measured in impulsive events. The lower particle intensities and energies typically observed in impulsive events are also characterized by enhanced $^{3}$He abundances \citep{balasubrahmanyan1974solar}, elevated electron to proton ratios, and high charge states of heavy ions \citep{reames1999particle}. Diffusive shock acceleration \citep[DSA; e.g.,][]{baring1997diffusive} is commonly believed to be the main acceleration mechanism responsible for the particle energization in gradual events. Most gradual SEP events are associated with the occurrence of fast CMEs, whose shocks can form in the solar corona and propagate through the heliosphere. If the conditions are appropriate, the CME shocks can continuously accelerate particles as they propagate from their formation in the low corona and as they move outward in interplanetary space. The precise location of the particle acceleration site by a propagating shock is uncertain. Observations of the early phases of CMEs \citep{balmaceda2022hyper} suggest that
the estimated speeds during the hyper-inflation phase (in which the CME undergoes a rapid lateral expansion) can be sufficiently high to generate shocks and to accelerate particles in the low corona.
In order to distinguish whether particles are accelerated close to the Sun, in the outer corona, or in interplanetary space, the study of a SEP event associated with a slow CME might help to localize this acceleration site, since such weak CMEs are not expected to continuously drive strong shocks in interplanetary space.
\citet{long2021localized} presented observations of energetic electron acceleration through measurements of Type III radio emissions associated with a very weak shock in the corona with an Alfv\'en Mach number of $\sim 1.008-1.013$ and shock speed of $\sim 400 - 600 $ $\rm km\;s^{-1}$. Observing SEP ions from weak CMEs at 1 au is difficult because weak CMEs are not expected to drive strong shocks able to accelerate particles with enough energy and intensity to be observed above the background of particle instruments and of galactic cosmic ray intensities.

Although many spacecraft have expanded our ability to probe the properties of SEPs through in-situ measurements and remote-sensing observations, reliable models to predict SEP radiation hazards are still lacking. Physics-based numerical models for the propagation and acceleration of SEPs from their source to Earth can help the prediction of SEPs. Over the years, several simulation tools have been developed to study the propagation of SEPs \citep[e.g.,][]{heras1992influence,heras1995three,Kallenrode1993, Bieberetal1994, Droge1994, NgReames1994, Ruffolo1995, KallenrodeWibberenz1997,lario1998energetic, zank2000particle, giacalone2000small, Ngetal2003, Riceetal2003, Lietal2003, Lee2005, qin2006effect, zhang2009propagation, Drogeetal2010, Luhmannetal2010, Kozarevetal2013, Marshetal2015, Huetal2017, ZhangZhao2017}. Closely relevant to this paper are the works by \citet{zhang2009propagation} and  \citet{droge2010anisotropic} that modeled the propagation of SEPs by solving the Fokker-Planck transport equation with stochastic processes in a three-dimensional (3D) interplanetary magnetic field, where the Parker model of the interplanetary medium is used. The Parker model describes reasonably well the undisturbed magnetic field and solar wind plasma in interplanetary space, but the magnetic field in the solar corona is far more complicated and dynamic, making it difficult to use a fixed configuration to describe it adequately. SEP production and propagation through the solar corona cannot be modeled without a model of magnetic field and plasma in this region. Recently, a data-driven coronal magnetic field configuration has been adopted by applying a Potential-Field Source-Surface (PFSS) model to synoptic magnetogram measurements \citep{zhang2017precipitation} obtained from a number of magnetographs such as NSO/GONG (\url{https://gong.nso.edu/data/magmap/archive.html}). On the other hand,  in this model and most previous models, SEP production by CME shocks is an ad hoc input. For example, in some models energetic particles were injected with an assumed energy spectrum at a fixed radial distance in the corona. However, the production of SEPs by a propagating CME-driven shock is far more complicated. First, the CME shock can continuously accelerate particles at different radial distances as it moves away from the Sun. Second, the properties of the CME shock vary with time, radial distance and along its front. Third, the shocks driven by CMEs vary significantly from event to event, and the inclusion of the shock as a mobile source of particles should be based as much as possible on observations. A SEP model for space weather predictions should consider the propagation of the CME shock through the corona and interplanetary medium as realistically as possible. In this paper, the SEP model calculation \citep{zhang2009propagation} is extended by injecting source particles at the location of the shock front reconstructed from coronagraph observations using an ellipsoid model \citep{kwon2014new} that allows us to capture realistic CME shock conditions. We also include a SEP seed injection model to determine particle intensity level from the input of shock properties. 

Our newly developed SEP model is applied to an SEP event that occurred on 2020 May 29 and was observed by both \textit{Parker Solar Probe (PSP)}, a spacecraft getting closer to the Sun than ever before, and by \textit{STEREO-A} in a $\sim$1 au orbit around the Sun. We compute the time profiles of the particle intensities observed by both spacecraft, derive the event-integrated particle energy spectrum, and estimate the path length of the magnetic field line experienced by the first arriving ions. We found that the simulation results are well correlated with observations. The flux of 2.2 MeV protons from both the simulation and the \textit{PSP} measurements are in the same order of magnitude with the peak value of 0.32 (cm$^{2}$ s sr MeV)$^{-1}$
and 0.15 (cm$^{2}$ s sr MeV)$^{-1}$, respectively.  The modeled flux of 5.0 MeV protons has a peak value  around $3.6\times 10^{-3}$ (cm$^{2}$ s sr MeV)$^{-1}$, which is between the  peak value of the 1.8\textendash3.6 MeV  and 4.0-6.0 MeV proton intensities measured by the Low-Energy Telescope \citep[LET;][]{mewaldt2008low} on \textit{STEREO-A}. The estimated spectral index is 2.08 from the simulation, which is consistent with the slope 2.18 obtained from \textit{PSP} measurements. We obtain an estimated equivalent path length of 0.70 au, which is lightly longer than the 0.625 au estimated by \citet{chhiber2021magnetic} using PSP observations. The SEP event could be explained by the acceleration of particles by a weak shock in the low corona that did not establish direct magnetic connection with the spacecraft via nominal Parker spiral interplanetary magnetic field lines.

The paper is organized as follows. Section \ref{sec:model} describes the simulation model. The methodology to solve the SEP acceleration and propagation equation in an arbitrary 3D magnetic field geometry to derive the SEP time-intensity profile at a given heliospheric location is presented. The simulation results are described in Section 3. Finally, Section \ref{sec:disc} presents a summary and discussion.

\section{Model Description} \label{sec:model}

The simulation tool used in this paper was initially developed by \citet{zhang2009propagation} and modified recently to make an efficient calculation of SEP flux using data-driven models of the corona, heliosphere, and CME shock. The governing transport equation of particle distribution function $f(t,\mathbf{x},p,\mu)$ as a function of time $t$, position $\mathbf{x}$, momentum $p$, and pitch-angle cosine $\mu$ can be written as \citep[e.g.,][]{zhang2009propagation}:
\begin{eqnarray}\label{eqn:trans} 
\frac{\partial f}{\partial t} - \nabla \cdot \bm{\kappa}_{\perp} \cdot \nabla f+\left(v \mu \hat{\mathbf{b}}+\mathbf{V}+\mathbf{V}_{d}\right) \cdot \nabla f -\frac{\partial}{\partial \mu} D_{\mu \mu} \frac{\partial f}{\partial \mu}+\frac{d \mu}{d t} \frac{\partial f}{\partial \mu}+\frac{d p}{d t} \frac{\partial f}{\partial p} = Q_0,
\end{eqnarray}
where the terms on the left-hand side come from particle transport mechanisms: cross-field spatial
diffusion with a tensor $\bm{\kappa}_{\perp}$, streaming along the ambient magnetic field or average magnetic field direction $\hat{\mathbf{b}}$ with particle speed $v$ and
pitch-angle cosine $\mu$, convection with the background plasma velocity
$\mathbf{V}$, particle gradient\slash curvature drift $\mathbf{V}_d$, pitch-angle diffusion
with a coefficient $D_{\mu \mu}$, focusing $\frac{d \mu}{d t}$, and adiabatic cooling
$\frac{d p}{d t}$.  On the right hand side of equation \ref{eqn:trans}, the term $Q_{0}$ represents the source rate of particles from a seed population at energies much lower than those of the SEPs measured during a SEP event (see description below).
 
Under the adiabatic approximation, the drift velocity, focusing rate, and cooling rate  may be calculated from the
ambient magnetic field $\mathbf{B}=B \hat{\mathbf{b}}$ and plasma velocity $\mathbf{V}$ through
\begin{eqnarray}\label{eqn:vd}
\mathbf{V}_{d}&=& \frac{c p v}{q B}\left\{\frac{1-\mu^{2}}{2} \frac{\mathbf{B} \times \nabla {B}}{B^{2}}+\mu^{2} \frac{\mathbf{B} \times[(\mathbf{B} \cdot \nabla) \mathbf{B}]}{B^{3}}\right. \nonumber \\ &&\left.+\frac{1-\mu^{2}}{2} \frac{\mathbf{B}(\mathbf{B} \cdot \nabla \times \mathbf{B})}{B^{3}}\right\},\\
\frac{d \mu}{d t} &=&-\frac{\left(1-\mu^{2}\right) v}{2} \hat{\mathbf{b}} \cdot \nabla \ln B \nonumber \\ &&+\frac{\mu\left(1-\mu^{2}\right)}{2} 
\times(\nabla \cdot \mathbf{V}-3 \hat{\mathbf{b}} \hat{\mathbf{b}}: \nabla \mathbf{V}) \nonumber \\ && -\frac{\left(1-\mu^{2}\right) m}{p}(\mathbf{V} \cdot \nabla \mathbf{V}) \cdot \hat{\mathbf{b}}, \\
\frac{d p}{d t}=&&-\left[\frac{1-\mu^{2}}{2}(\nabla \cdot \mathbf{V}-\hat{\mathbf{b}} \hat{\mathbf{b}}: \nabla \mathbf{V})+\mu^{2} \hat{\mathbf{b}} \hat{\mathbf{b}}: \nabla \mathbf{V}\right] \nonumber \\
&& \times p-\frac{\mu p}{v}(\mathbf{V} \cdot \nabla \mathbf{V}) \cdot \hat{\mathbf{b}},
\end{eqnarray}
where $q$ and $m$ are the charge and mass of the particles, respectively.
The formulas for the terms in the first-order partial derivatives can be found in
many previous publications \citep{northrop1963adiabatic,isenberg1997hemispherical,qin2004interplanetary,qin2006effect}. 
The second-order partial derivative terms represent the effects of magnetic field
turbulence. The equation is truncated up to the diffusion term
as approximated in the standard quasi-linear theory. All the diffusion terms related to $p$ are neglected, considering that the propagation speed of magnetic field turbulence, typically the Alfv\'en speed or fast-mode MHD wave speed, is much less than the speed of particles,  and stochastic particle momentum change by electric field fluctuations in the turbulence is much slower than the adiabatic cooling by the background solar wind plasma. If we assume that phases of magnetic field turbulence with a steeply decreasing power spectrum at different wavelengths are completely random or independent, pitch-angle scattering and cross-field spatial diffusion become uncorrelated, yielding zero off-diagonal diffusion elements in the diffusion tensor \citep{jokipii1966cosmic}.

Like the Parker transport equation, the focus transport equation (\ref{eqn:trans}) can be applicable to shock acceleration \citep{leRouxWebb2012, zuoetal2013a, zuoetal2013b}. It is accurate when the particle velocities are much greater than the shock speed \citep{zhang2009propagation}. The focus transport equation can allow the particle distribution function to be very anisotropic, which is required for describing SEPs near the Sun or in the early phase of an SEP event. Protons above MeV energies accelerated by CME shocks in the low corona discussed in this paper are suitable for the focus transport equation (\ref{eqn:trans}). 

We seek the solution to the transport equation ({\ref{eqn:trans}) to get the distribution of particles as a function of time, energy and pitch angle at any particular heliospheric location. The equation is a time-dependent 5-dimensional second-order (Fokker-Planck) partial differential equation in the phase space. Typical finite difference or finite element methods become impractical for this high dimensional application. We use time-backward stochastic differential equations derived from the left-hand side of the focus transport equation (\ref{eqn:trans}) to describe the motion of the particle guiding center and momentum \citep{gardiner1985handbook, zhang2009propagation}
\begin{eqnarray}\label{eqn:bk}
d \mathbf{x}(s)&=&\sqrt{2 \kappa_{\perp}} \cdot d \mathbf{w}(s)+\left(\nabla \cdot \boldsymbol{\kappa}_{\perp}-v \mu(s) \hat{\mathbf{b}}-\mathbf{V}-\mathbf{V}_d\right) d s,\nonumber\\ \\
d \mu(s)&=&\left[-\frac{d \mu}{d t}+\frac{\partial D_{\mu \mu}}{\partial \mu}\right] d s+\sqrt{2 D_{\mu \mu}} d w(s),\\
d p(s)&=&-\frac{d p}{d t} d s, \label{eqn:bk3}
\end{eqnarray}
where $d w(s)$ is a Wiener process as a function of $s$, which is the time running backward. $d w(s)$ can be generated by random numbers from a Gaussian distribution with a standard deviation of $\sqrt{ds}$.
The simulation of stochastic processes starts at the location ${\bf x}$, pitch-angle $\mu$, momentum $p$, and time $t$ where the solution to the particle distribution function is sought, i.e., ${\bf x}(0) = {\bf x}$, $\mu(0) = \mu$, and $p(0) = p$ at initial backward time $s = 0$ starting at the observation time $t$. 

An exact solution to the transport equation (\ref{eqn:trans}) for any location, momentum, pitch angle cosine and time can be written as \citep{Freidlin1985}
\begin{eqnarray}\label{solution}
f(t,{\bf x},p,\mu) = \left < \int_0^t Q_0(t-s, {\bf x}(s),p(s),\mu(s)) ds \right > + \left < f_b(t-s_e, {\bf x}_e,p_e,\mu_e) \right >
\end{eqnarray}
where $\left < \right > $ denotes the expectation of what is inside and $f_b(t-s_e, {\bf x}_e,p_e,\mu_e)$ is the boundary or initial value 
of the distribution function when the stochastic processes hit a boundary or the initial time (the subindex $e$ refers to the first exit of the simulation). For our simulation of SEPs produced by CME shocks, $f_b = 0$ if we choose the initial time before the solar eruption and set the inner boundary at the solar surface and outer boundary at a large enough radial distance. Therefore, the solution to the transport equation (\ref{eqn:trans}) is the expectation of the source injection rate integrated over time along backward stochastic trajectories. This is a major difference with respect to the work by \cite{zhang2009propagation} where a boundary condition at a given inner distance in the upper corona was assigned as a point for the injection of SEPs.  We now run stochastic trajectories backward in time from the location, energy, and pitch angle where we want to calculate the particle intensity until CME initiation. Trajectories that encounter particle sources at shock crossings will contribute to the average. Many simulated trajectories do not encounter the shock, so we design a scheme to drive the simulated stochastic trajectories toward the shock by introducing an artificial drift. It can be achieved by substituting
$f$ with $(1+a\mu_{r})f$ into the transport equation, where $a$ is a tuning parameter and $\mu_r$ is the cosine of pitch angle to the magnetic field line outward from the Sun. The driven-up probability towards particle source at the shock is compensated by an exponential decay or killing term that weighs on the average. See \cite{zhang2009propagation} for a detailed description of this methodology. We typically simulate $10^4$ to $10^5$ trajectories until a statistically significant number of useful trajectories or a small enough error bar (i.e., a relative error bar $\delta f / f$ of less than 10 \%) for the expected value has been achieved. The value of $a$ does not affect the computation result once enough statistics are achieved, but it does influence how fast the result can converge in the computer simulation.  

The seed particle source of SEPs $Q_0$ is typically of energies much lower than the measured SEPs we intend to simulate. Contribution of the SEP seed particles to the averaging of the distribution function $f$ in equation (\ref{solution}) as a function of momentum $p$ is through particle acceleration at the CME shock with the term containing $dp/dt$ in the transport equation (\ref{eqn:trans}) or stochastic differential equation (\ref{eqn:bk3}). The detailed processes of diffusive shock acceleration must be simulated in order to correctly capture the amount of acceleration and seed particle source injection. Acceleration of SEPs from the seed particles occurs on a small scale near the shock ramp. Simulation of such acceleration processes takes a huge amount of computation time, thus becoming impractical for large-scale simulation of SEP production and transport. 

We take an alternative approach to incorporate diffusive shock acceleration in this model. We note that the steady-state DSA solution provides a momentum distribution of particles given by a power law with a slope $\gamma_s = 3R/(R-1)$ determined only by the shock compression ratio $R$ up to a cut-off momentum ($p_c$) independent of the particle diffusion coefficient \citep[e.g.,][]{Drury1983} and the large-scale shock geometry. It is unlikely that SEP transport on the large-scale heliospheric magnetic field will affect the local shock acceleration of particles below the cut-off momentum. Therefore, the particle distribution function at the shock is known as long as we know how many total seed particles have been injected at the shock location. We can move the term of particle acceleration at the shock front to combine with the seed source rate to get a new accelerated SEP injection rate as
\begin{eqnarray}\label{eqn:source1}
Q =Q_0 + \frac{d p}{d t}_{sh} \frac{\partial f_{sh}(p<p_c) }{\partial p},
\end{eqnarray}
where 
\begin{equation}\label{shockspec}
f_{sh} = f_{sh0} \left(  \frac{p}{p_{inj}}\right) ^{-\gamma_s} {\rm ~for~} p<p_c. 
\end{equation}
and $p_{c}$ and $p_{inj}$ will be determined as described below (see equations (\ref{acctime}) and (\ref{totalsource})). 
Once the shock acceleration term is combined with the source term, the gain of particle momentum during the shock passage is no longer included in the stochastic differential equation according to the correspondence between the Fokker-Planck equation and stochastic differential equation.

Note that the acceleration or cooling term in the transport equation away from the shock front location and above $p_c$ is still left in the particle transport calculation, and their effects on particle momentum gain or loss still are followed by the simulation of stochastic particle transport processes. 
Because $\frac{dp}{dt}_{sh}$ is a $\delta$-function on the shock surface, where the plasma and magnetic field properties are discontinuous, the rate of momentum change $dp/dt$ is ambiguous primarily due to the discontinuity in the magnetic field direction relative to the shock normal. We average the shock SEP injection over all particle pitch angles to avoid such ambiguity, assuming that the particle distribution at the shock front is isotropic. We have compared these results with a calculation using an anisotropic acceleration term, and we found that the difference is minimal, probably because the particles do cross the shock in pitch angles very close to an isotropic distribution. The isotropic assumption is justified because of the expected enhanced turbulence in the vicinity of a shock. So the accelerated SEP source rate can be written as follows:
\begin{eqnarray}\label{eqn:source2} 
Q = Q_0+ \frac{1}{3} \left(V_{2}-V_{1}\right) \delta\left(\mathbf{x}-\mathbf{x}_{s h}\right) p \frac{\partial f_{s h}}{\partial p},
\end{eqnarray}
where $V_1$ and $V_2$ are plasma speeds relative to the shock upstream and downstream, respectively, and $x_{sh}$ is the instantaneous shock location. $Q$ replaces the $Q_0$ in equation (\ref{solution}) in the calculation of the solution to the particle distribution function. The integration of the delta function over time $\delta\left(\mathbf{x}-\mathbf{x}_{s h}\right) ds$ is called local time. We use Tanaka's formula and Ito stochastic calculus up to the second order to calculate the differential local time for each shock crossing \citep[e.g.,][]{bjork2015pedestrian}. The majority of particle acceleration takes place at the shock, and its effect has been represented by the addition of the accelerated SEP source injection. Without shock acceleration along the simulated trajectories, the particles starting at SEP with energies above 1 MeV will never decrease their energy low enough to have a direct contribution from the seed particle population of typically a few keV. Essentially, $Q_0$ can be considered zero, but the seed particles contribute indirectly through the injection of accelerated SEPs at the shock, which is constrained by a power-law distribution according to the theory of diffusive shock acceleration. In this way, we can speed up the computation and incorporate the shock acceleration without simulating the entire acceleration processes. It is correct as long as acceleration obeys the result of diffusive shock acceleration theory. 

The cutoff momentum of the shock power-law distribution $p_c$ can be determined by the amount of time available for diffusive shock acceleration $t_{acc}$ \citep{Drury1983}
\begin{equation}\label{acctime}
t_{acc}=\int_{p_{inj}}^{p_c} \frac{3}{V_1-V_2} \left [ \frac{\kappa_1}{V_1}+\frac{\kappa_2}{V_2} \right] \frac{dp}{p}
\end{equation}
where $p_{inj} = m v_{inj}$ (see discussion on particle injection speed later) is the momentum of injected seed particles, $\kappa_1$ and $\kappa_2$ are the particle diffusion coefficients upstream and downstream of the shock, respectively. Because typically $\kappa_2 \ll \kappa_1$, the upstream condition essentially determines the acceleration time. We choose the Bohm limit for it or $\kappa_1 = v p /(3 q B_1)$, where $v$ is the particle speed, $q$ particle charge and $B_1$ is upstream magnetic field field strength. Because of the increasing diffusion with momentum, the acceleration time has little dependence on $p_{inj}$ ($\ll p_c$). The cutoff momentum $p_c$ can be determined through equation (\ref{acctime}) by setting $t_{acc}$ to be equal to the minimum between the shock lifetime $t$ since the onset of the CME eruption and the adiabatic particle energy loss time in the background solar wind plasma $t_{loss} = 3 (\nabla \cdot {\bf V})^{-1}$, i.e., $t_{acc} = {\rm min}(t, t_{loss})$.

Coronal magnetic field covering the region from 1 $\rm R_S$ to 2.5 $\rm R_S$ is adopted into the SEP model by applying the PFSS model to synoptic magnetograms observations (the first one after the eruption) \citep{zhang2017precipitation}, where $\rm R_S$ is the solar radius. The results do not change much with time-dependent synoptic magnetograms \citep{Zhao_2018}. The computation of SEP transport is done in the reference frame corotating with the Sun, where the coronal magnetic field is stationary. Beyond 2.5 $\rm R_S$, a Parker model of the heliospheric magnetic field with an empirical solar wind speed and density profile \citep{leblanc1998tracing} is used. In the corotation frame, the tangential component of the plasma velocity is $-\Omega_S (r-2.5 R_S)$, with $\Omega_S$ being the angular rotation speed of the Sun. It means that the solar wind begins to lose corotation at 2.5 $R_S$ and becomes nonrotating at large enough radial distances. The computation domain covers the inner boundary at 1 $\rm R_s$ to an outer boundary at a radial distance of 20 au, far enough that it does not affect the calculation result in the inner heliosphere unless an abnormally large mean free path is used. Both the inner and outer boundaries are absorptive boundaries, where the SEP distribution function is set to zero. 

The sources of accelerated SEPs comove with the CME shock. The location, shape, and time propagation of the CME are taken from an ellipsoid model developed by \citet{kwon2014new}. A 3D CME shock surface is reconstructed using EUV and white-light coronagraph images from instruments on \textit{STEREO}, \textit{SDO}, and \textit{SOHO}, which cover a radial range from a few $\rm R_S$ to tens of $\rm R_S$. Depending on the solar eruption, each event typically contains several time frames from multiple vantage points that allow an identification of the large-scale CME shock geometry and its propagation through the solar corona. Beyond the last frame, when the CME shock extends out of the field of view of the coronagraphs, a friction CME shock propagation model is adopted to model the slowdown of CME shock. The radial distance of the shock front at the nose $r_{shf}$ slows down according to 
\begin{equation}\label{frictionmodel}
r_{shf}=r_{shf0}+V_{sw}\left(t-t_0\right)+\frac{\left(V_{shf0}-V_{sw}\right)}{b}\left(1-e^{-b \left(t-t_0\right)}\right)
\end{equation}
where $r_{shf0}$ and $V_{shf0}$ are the initial radial distance and speed of the shock front at the nose as determined by fits to coronagraph observations, respectively, and  $V_{sw}$ is the solar wind speed, $b$ in the dimension of $1/t$ is the parameter to measure the rate of shock slowdown.
We use a parameter $t_{1au}$ to control the time when CME shocks pass 1 au and estimate the deceleration rate of the CME shock. We insert the ellipsoid shock surface and its time evolution into the coronal and heliospheric magnetic field and plasma configuration to derive the shock properties at any point on the shock surface. Relevant parameters, such as shock speed relative to the plasma, shock normal, upstream magnetic obliquity, Alfv\'en Mach number, and sonic Mach number, are fed into the MHD shock adiabatic equation \citep[e.g.,][]{Thompson1962, Kabin2001JPlPh..66..259K} to numerically derive the fast-mode plasma compression ratio at the shock, which further yields the downstream plasma density, velocity, magnetic field, and plasma thermal speed. The shock compression ratio is used to determine the slope of the SEP power-law spectrum. We assume that particles are injected at the shock from the thermal tail of solar wind ions. The shock can substantially heat solar wind ions to become sub-magnetosonic after the shock crossing. Immediately downstream of the shock, the thermal tail particles have  high enough energies to overcome convection away from the shock, and they are more likely to become the seed particles for diffusive shock acceleration. In this simulation, we choose a characteristic particle injection speed to be 2.5 times the shock speed, $v_{inj}= 2.5 V_{sh}$. The amount of injected seed particles can be related to the Maxwellian velocity distribution of downstream solar wind ions so that
\begin{equation}\label{totalsource}
f_{sh0} = \frac{n_{sw2}}{(4\pi v_{th2}^2)^{3/2}} exp\left(-\frac{v^2}{v_{th2}^2} \right)
\end{equation}
where $n_{sw2}$ and $v_{th2}$ are the downstream solar wind density and the thermal speed.
Because the choice of the injection speed sits in a Maxwellian tail, the number of total injected particles is sensitive to $v_{inj}$. We found that a $v_{inj}$ between $2.3 -2.7$ $V_{sh}$ can generally produce a good fit to observations. Therefore, we set $v_{inj}= 2.5 V_{sh}$ in this simulation. In addition, the code can handle arbitrary sources of seed particles. If a particular suprathermal population is injected, we can add the total number of injected particles to equation (\ref{totalsource}). 

Since the ellipsoid fitted shock is inserted on the plasma and magnetic field configuration obtained from the PFSS model, the downstream plasma and magnetic field distribution inside the shock ellipsoid is not consistent with the shock jump condition. The calculation of particle acceleration is not entirely correct unless we modify the downstream magnetic field and plasma. This requires an input of a time-dependent plasma and magnetic field model. However, the effect of shock acceleration has been replaced by the injection of accelerated SEPs with a power-law spectrum determined by the shock compression ratio up to momentum $p_c$, so we do not have to correct for the change of plasma and magnetic field due to the CME shock propagation. For particles above $p_c$, the result still somewhat relies on calculating the shock acceleration process from a momentum below $p_c$, which is not an issue for most of our applications in this paper. 

Our model also requires an input of particle transport coefficients, such as pitch angle diffusion coefficient $D_{\mu\mu}$ and spatial diffusion perpendicular to the magnetic field $\kappa_\bot$. No direct measurements of these transport coefficients are possible. The value can only be estimated through the magnetic field turbulence spectrum with aid from a theory of particle transport coefficient. Alternatively, they can be treated as free parameters to fit observations. In this simulation, we follow an approach we have adopted in a previous work  \citep[e.g.,][]{ZhangZhao2017}. We assign 
\begin{equation}
D_{\mu\mu} = D_0(r) p^{q-2} (1-\mu^2) (|\mu|^{q-1} +h_0)
\end{equation}
with $q=5/3$, and $h_0=0.2$. The additional parameter $h_0$ is to phenomenologically describe the enhancement of scattering through $\mu = 0$ by either nonresonant scattering or nonlinear
effects.  $D_0(r)$ is chosen such that the radial mean free path is constant at 10.0 $\rm R_S$ or 0.0465 au for particles of 1 GV rigidity. This formula is based on the quasilinear theory of particle scattering by magnetic field fluctuations with a Kolmogorov spectrum, plus a non-linear correction at the 90$^\circ$ pitch angle. 
The perpendicular diffusion is assumed to be mainly driven by field line random walk started at the bottom of the solar corona. A derivation by \cite{ZhangZhao2017} yielded
\begin{equation}
\kappa_\bot = \frac{v}{2V} k \kappa_{gd0} \frac{B_0}{B} 
\end{equation}
where $\kappa_{gd0} = 3.4 \times 10^{13}$ cm$^2$ s$^{-1}$ is the diffusion coefficient in the photosphere estimated from a typical speed of supergranular motion, $v/V$ is the ratio of particle to solar wind plasma speed, and $B_0/B$ is the expansion of magnetic field flux tube from the solar surface. A factor $k$ is inserted to tune down the transmission of field line diffusion from the photosphere to the corona. In this simulation we set $k=0.074$. 

\section{Results} \label{sec:res}
We now apply our model calculation to the 2020 May 29 SEP event, one of the six SEP events observed during 2020 May 22$-$June 1 as described by \citet{chhiber2021magnetic} and \citet{cohen2021parker}  \citep[see also][]{zhuang2022widespread}. We choose to simulate this SEP event because the associated CME-driven shock can be reconstructed from coronagraph measurements. On 2020 May 29, \textit{PSP} was at a heliocentric radial distance of 0.33 au (71.0 $\rm R_S$). The involved Active Region (AR) was AR12764 at N34 in the northern hemisphere. A slow CME with a plane-of-sky speed estimated as 337  $\rm km\;s^{-1}$ at heights below 6 $\rm R_s$ has been reported in the SOHO/LASCO CME catalog \citep[\url{https://cdaw.gsfc.nasa.gov/CME_list/index.html};][]{2009EM&P..104..295G}.

The locations of \textit{PSP}, \textit{Earth}, and \textit{STEREO-A} at 08:04 UT on 2020 May 29 are labeled by the letters P, E, and A in Figure \ref{fig:CME}(a), respectively, where the Sun is at the center. Sample magnetic field lines are shown as spiral curves, with the field lines passing through \textit{PSP} shown as red curves. The blue arrow denotes the direction into which the CME headed (i.e., $\sim92^{\circ}$ east in longitude from the Earth-Sun line). The reconstructed CME shock viewed from two different angles is indicated by the green surface in Figures \ref{fig:CME}(b) and  \ref{fig:CME}(c), where the white sphere indicates the solar surface.  \textit{STEREO-A} was located 72$^{\circ}$ in longitude east of Earth and at a heliocentric radial distance  of 0.96 au (the 1 au distance is indicated by the dashed circle in Figure \ref{fig:CME}(a)). \textit{PSP} was $151^{\circ}$ in longitude ahead of Earth. None of the spacecraft, \textit{PSP}, \textit{STEREO-A} and \textit{SOHO}  (this latter near the Earth-Sun Lagrangian L1 point) established magnetic connection with the reconstructed CME shock via nominal Parker spiral  interplanetary magnetic field lines. 

\begin{figure}
        \epsscale{0.9}
	\plotone{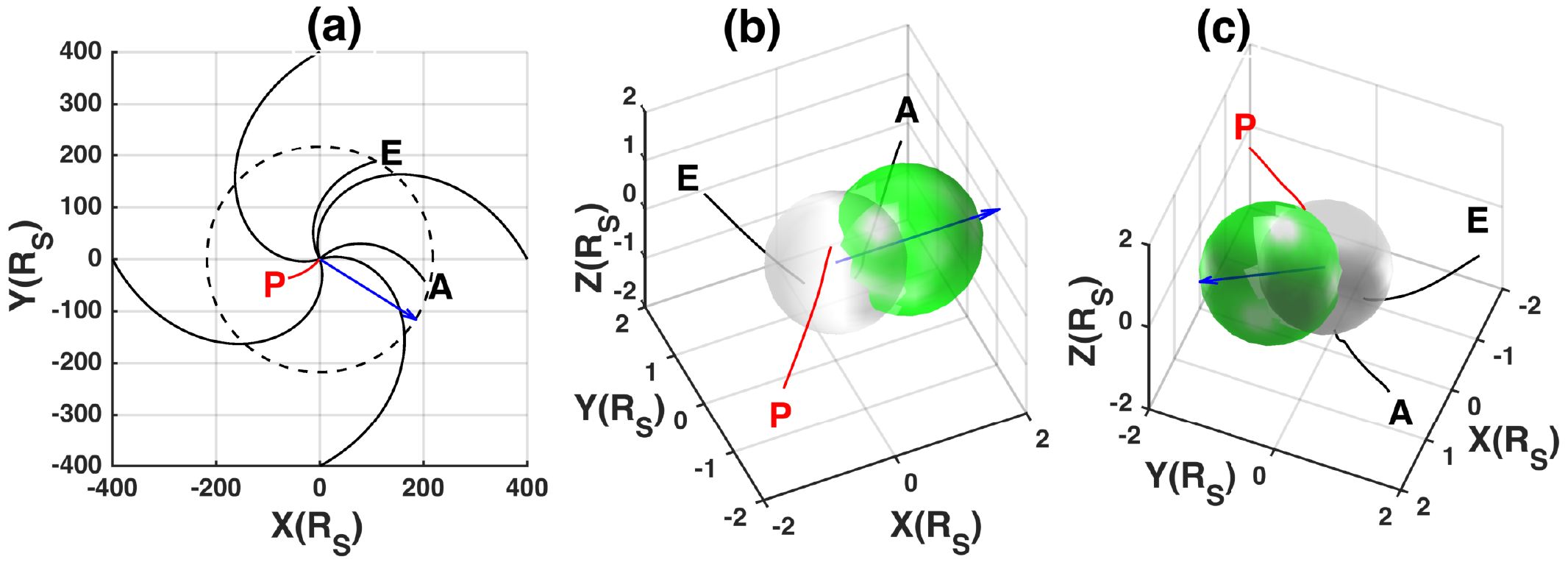}
	\caption{(a) Locations of \textit{PSP} (P), \textit{Earth} (E) and \textit{STEREO-A} (A) in the equatorial plane as seen from the north. The blue arrow denotes the moving direction of the CME shock; The red and black curves are the Parker spiral and coronal magnetic field lines that connect to the spacecraft. A shift ($60^{\circ}$) of the heliocentric Earth equatorial (HEEQ) coordinate system in longitude is used. Earth is at $60^{\circ}$ in longitude. The panels (b) and (c) show the reconstructed 3D CME shock marked by the green surface from two different view angles.  
		\label{fig:CME}}
\end{figure}

The CME on 2020 May 29 had a very slow speed. Figure {\ref{fig:skf} shows the time evolution of the properties of the wavefront at the leading point along the direction indicated by the blue arrow in Figure \ref{fig:CME}(b). Also shown are the time variation of the radial distance of the shock leading point ($r_{shf}$), the wavefront speed relative to the upstream plasma ($u_1$), the Alfv\'en speed ($V_A$), Alfv\'en Mach number ($Ma_A=u_1/V_A$), fast magnetosonic Mach number ($Ma_{MS}$) and the shock compression ratio ($R$). The shock only exists in the radial distance range of around 1.7 - 2.5 $\rm R_S$ from 07:38 UT to 08:10 UT. Below 1.7 R$_S$, the wavefront is slower than the Alfv\'en speed. Starting from the radial distance of 2.5 R$_S$ or time 08:10 UT, the solar wind speed has increased enough so that the wavefront speed relative to the plasma is less than Alfv\'en speed. The shock cannot form below $Ma_A=1$, and the compression ratio is set to 1. During the limited time period of the shock existence, the maximum value of the compression ratio is $\sim$1.6. Since the CME shock only survives up to 8:10 UT below 2.5 $\rm R_s$, the assumption about how the shock propagates or slows down at later time does not affect our result. Note that Figure  {\ref{fig:skf} is only for the leading point (or apex) of the reconstructed shock. It is possible that other regions of the shock front, e.g. propagating in a media with lower Alfv\'en speed, created a stronger shock. 

\begin{figure}
	\epsscale{0.95}
	\plotone{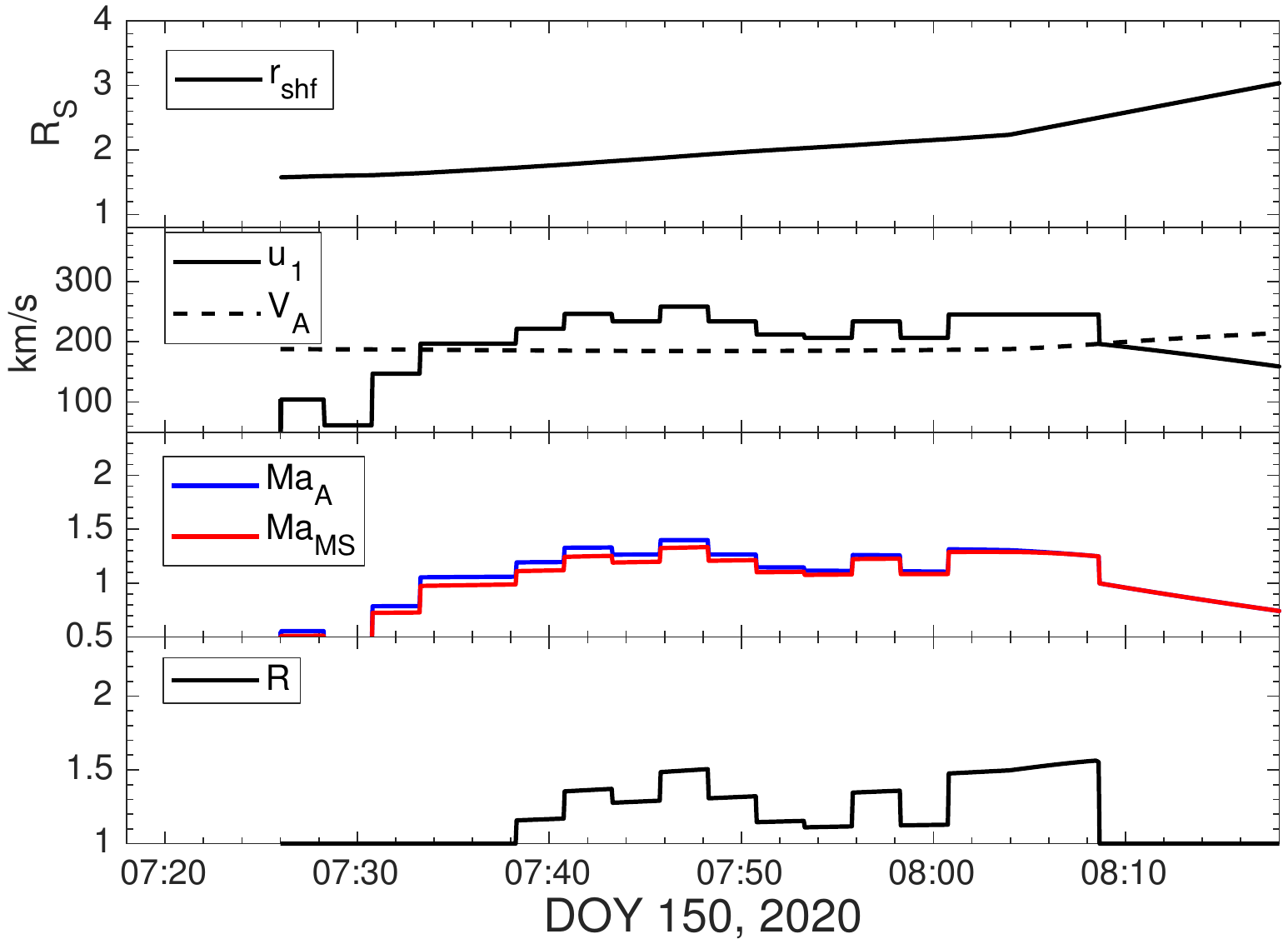}
	\caption{Time variation of (a) The radial distance; (b) the speed of wavefront ($u_1$) in the frame of solar wind, Alfv\'en speed ($V_A$); (c) Alfv\'en Mach number ($Ma_A$), magnetosonic Mach number ($Ma_{MS}$) and (d) compression ratio (R) throughout the existence of CME shock.   
		\label{fig:skf}}
\end{figure}

In spite of the lack of magnetic connection between \textit{PSP} and the CME shock, SEPs were seen to propagate to the location of \textit{PSP}. Assuming that the SEPs observed by \textit{PSP} were accelerated by the CME shock, one possibility for the particles to reach \textit{PSP} is through particle diffusion across the nominal average interplanetary magnetic field lines. There is also the possibility that the interplanetary magnetic field configuration differed from the nominal field configuration shown in Figure \ref{fig:CME}(a) as pointed out in Appendix A of \cite{zhuang2022widespread}, but this possibility will not be analyzed here because of the complexity and uncertainty of the exact magnetic field topology at the onset of the SEP event. Figure \ref{fig:fluxcomparison} shows the proton intensity-time profiles at two different energies observed by \textit{PSP} (open symbols) and obtained by our simulations (solid traces). Observational data are from the EPI-Hi instrument, which measures energetic protons and heavy ions  from $\sim$1 to $\sim$100 MeV\slash nuc \citep{mccomas2016integrated}. Overall, the evolution of the proton flux from the simulation is well correlated with the observations. The flux of 2.2 MeV protons from the simulation and the observation are in the same order of magnitude with the peak value of 0.32 (cm$^{2}$ s sr MeV)$^{-1}$
and 0.15 (cm$^{2}$ s sr MeV)$^{-1}$, respectively. Observationally, the onset of the 2.2 MeV proton intensity enhancement occurred around 08:30 UT, whereas the simulated 2.2 MeV proton intensities increased 1.5 hours later. Such a delay can probably be shortened by fine-tuning both parallel and perpendicular mean free paths, but here we do not perform a parameter search due to the demand of computation time. The simulated 12 MeV proton intensities display a similar behavior with respect to the observed intensities, with a delayed onset and a larger intensity. The peak intensities at both energies are within the observed values by a factor of 2 without fine-tuning the particle injection speed at CME shock. At the current stage of research, fine-tuning these parameters will not be necessary because, among other factors, the solar wind density of our plasma model can easily exceed a factor of 2 uncertainty. 
\begin{figure}
        \epsscale{0.65}
        	\plotone{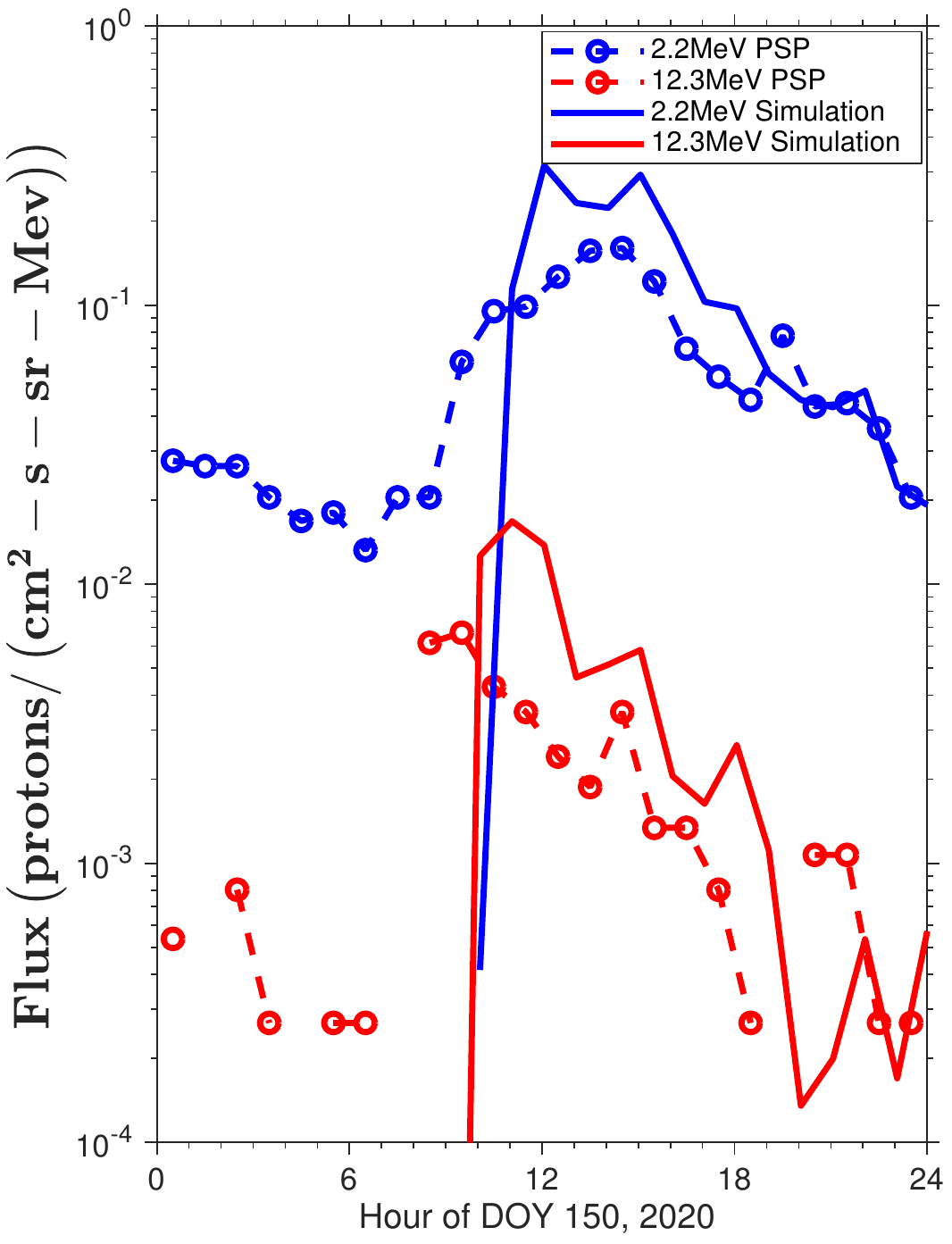}
	\caption{Time evolution of proton flux at 2.2 MeV (blue curves) and 12.3 MeV (red curves) from the simulation (solid traces) and \textit{PSP} observations (open symbols). Data are from the EPI-Hi instrument of \textit{PSP}. 
		\label{fig:fluxcomparison}}
\end{figure}

 Figure \ref{fig:powerlaw} shows the event-integrated energy spectra of the observed (blue) and simulated (red) SEPs. Observed and simulated spectra marked as diamonds and circles, respectively, are calculated by integrating the ion flux over the time period 07:30\textendash21:30 UT for all energetic particles below 20 MeV at \textit{PSP}. The blue (red) line shows the power-law fit to the spectrum from \textit{PSP} observation (simulation). The estimated spectral index is 2.08 from the simulation, which is consistent with the slope of 2.18 from the \textit{PSP} observation (2.31 for all energetic particles up to 30 MeV in \citet{cohen2021parker}). Although a seed population is needed to start the particle acceleration and determine the particle intensity level, according to the diffusive shock acceleration theory, the shape of the energy spectrum is not sensitive to seed particle injection. The close match between the observed and simulated spectra indicates that diffusive shock acceleration can explain the SEP production in this event. The estimated energy spectral index of 2.08 corresponds to a momentum spectral index $\gamma_{s}$=6.16. The compression ratio calculated from $\gamma_s = 3R/(R-1)$ should be 1.95, which is a little larger than the compression ratios obtained from the modeled shock structure as displayed in Figure \ref{fig:skf}. At a compression ratio of 1.60, the slope of particle distribution function at the shock is expected to be 8. The slightly higher observed event-integrated spectrum index may result from  energy-dependent particle propagation effects. 
  
 \begin{figure}
 	\epsscale{0.95}
	\plotone{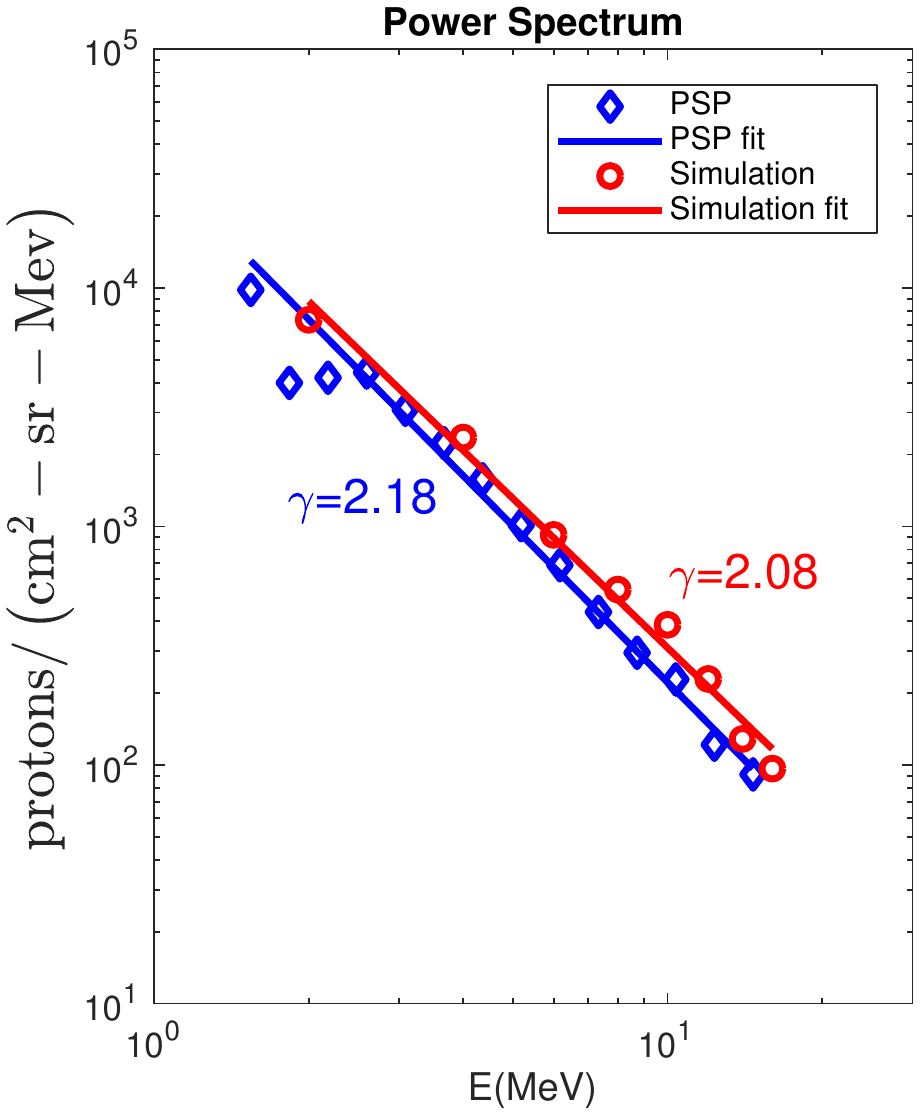}
 	\caption{Observed (diamonds) and simulated (circles) spectrum (fluence ($\rm {cm}^{-2}\;s^{-1}\;{MeV}^{-1}$)) of solar energetic particles at \textit{PSP}. The blue (red) line shows the power-law fit to the spectrum from \textit{PSP} observations (simulation). Estimated spectral index ($\gamma$) is 2.08 from the simulation and 2.18 from the observations. 
 		\label{fig:powerlaw}}
 \end{figure}

 We also estimate the onset time path length by the plot of reciprocal particle velocity 1\slash v versus onset time in Figure \ref{fig:pathlength}. The onset time is chosen as the time when the flux rises above 1$\%$ of the peak value. There is little change in the result of onset time when changing the criteria for the onset threshold slightly (a few percent). The linear relationship has allowed us to estimate an equivalent path length of 0.70 au, which is slightly larger than 0.625 au from the observations estimated by \citet{chhiber2021magnetic}. 
\begin{figure}
	\epsscale{0.65}
	\plotone{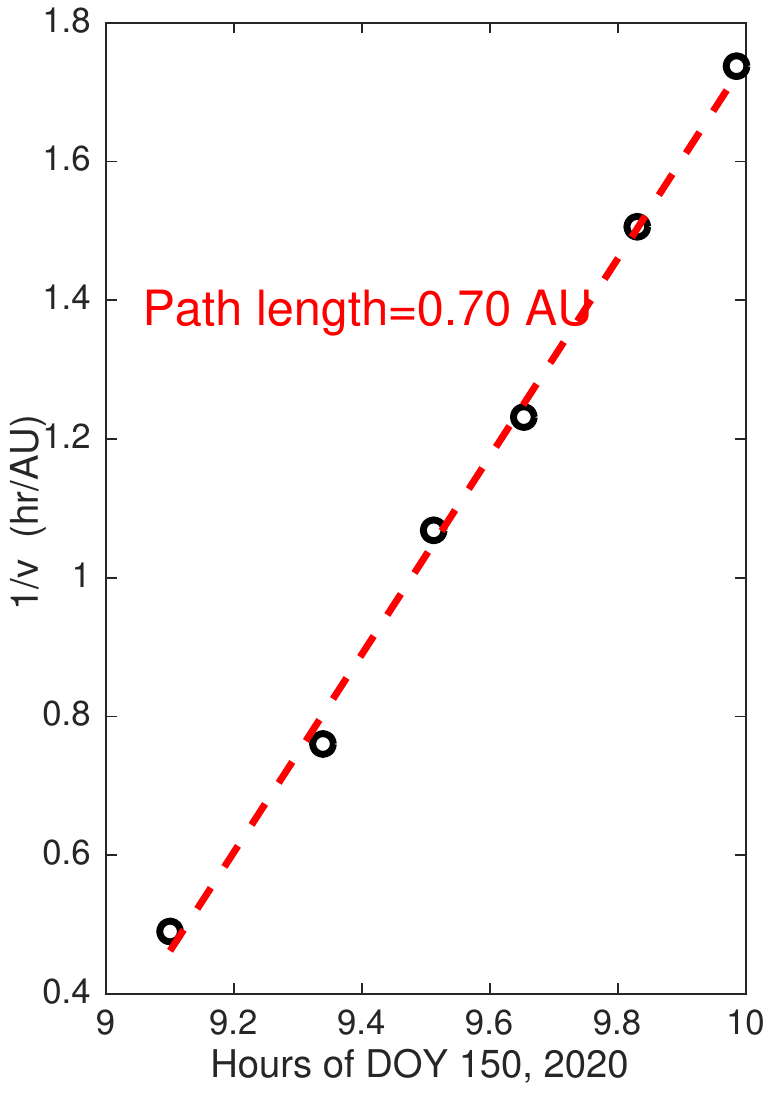}
	\caption{Onset time path length estimated by the plot of reciprocal particle velocity 1\slash v versus time from the simulation. The red dashed line shows a least chi-square linear fit line with a slope around 1.43 au$^{-1}$, which corresponds to a path length of $\sim$ 0.70 au.
		\label{fig:pathlength}}
\end{figure}
In our model, the length of the average interplanetary magnetic field line is approximately 0.33 au. The initial arriving particles experience an equivalent longer field line length. This is because in our model, the particle pitch angle relative to the averaged field has been scattered by magnetic field so that the average (over the course of the propagation from the source to \textit{PSP}) pitch angle cosine of the first arriving particle $\left < \mu \right >$ is approximately $0.47$, or $62^\circ$ pitch angle, even for the first arriving particles. It occurs because low-energy particles have small enough mean fee paths, and they have been sufficiently scattered before they arrive at \textit{PSP}. Without the consideration of intervening transient interplanetary structures \citep[such as the presence of an interplanetary CME; see][]{zhuang2022widespread}, the path length estimated by \citet{chhiber2021magnetic} assumes particles have an average pitch angle of 37$^\circ$ from the wiggling turbulent magnetic field lines. Although a very different picture of particle propagation, their result suggests the magnetic field lines, including the turbulent component, are corresponding to an average pitch angle of $25^\circ$. 

The SEP event is also seen by \textit{STEREO-A} around 10 hours later. Figure \ref{fig:CME} shows \textit{STEREO-A} does not connect to the shock. Unlike what is seen by \textit{PSP},  the flux of protons in low energy channels, such as in 1.11\textendash1.98 MeV energy range, does not rise above the background level of the measurements from the Solar Electron and Proton Telescope \citep[SEPT;][]{muller2008solar} on \textit{STEREO-A}  as shown by the blue dashed curve in Figure \ref{fig:stereoA}. To examine it, the time evolution of the simulated proton flux at \textit{STEREO-A} for 1.5 MeV (blue solid curve) and 5.0 MeV (red solid curve) is shown in Figure \ref{fig:stereoA}. The simulated particle intensity is obtained with the same model setup as it was used to simulate for the local of \textit{PSP}. The modeled flux of 5.0 MeV protons has a peak value  around $3.6\times 10^{-3}$ (cm$^{2}$ s sr MeV)$^{-1}$, which is between the  peak value of the 1.8\textendash3.6 MeV (green dashed curve) and 4.0-6.0 MeV (red dashed curve) proton intensities measured by the Low-Energy Telescope \citep[LET;][]{mewaldt2008low} on \textit{STEREO-A}, much larger than the background flux $\sim 1\times 10^{-4}$ (cm$^{2}$ s sr MeV)$^{-1}$ of 4.0\textendash6.0 MeV protons and $\sim 4.0\times 10^{-4}$ (cm$^{2}$ s sr MeV)$^{-1}$   of 1.8\textendash3.6 MeV protons. However, the predicted flux of 1.5 MeV protons from our simulation peaks around at  0.03 (cm$^{2}$ s sr MeV)$^{-1}$, which is smaller than the background ion flux $\sim 0.07$ (cm$^{2}$ s sr MeV)$^{-1}$ of 1.11\textendash1.98 MeV protons from \textit{STEREO-A}/SEPT. Thus, this explains why no clearly enhancement of 1.11\textendash1.98 MeV protons was seen by \textit{STEREO-A}/SEPT. Low-energy particles have too small mean free paths to reach 1 au before they dissipate in interplanetary space through adibatic cooling.  The peak intensity of high-energy protons at 5.0 MeV is comparable to the predicted peak level in our simulation without adjusting model parameter from the runs for \textit{PSP}. The rise of 4.0\textendash6.0 MeV proton flux appears later than the prediction, suggesting that the particle mean free path between 0.33 au and 1 au needs to be adjusted to a slightly lower value. 
\begin{figure}
	\epsscale{0.95}
	\plotone{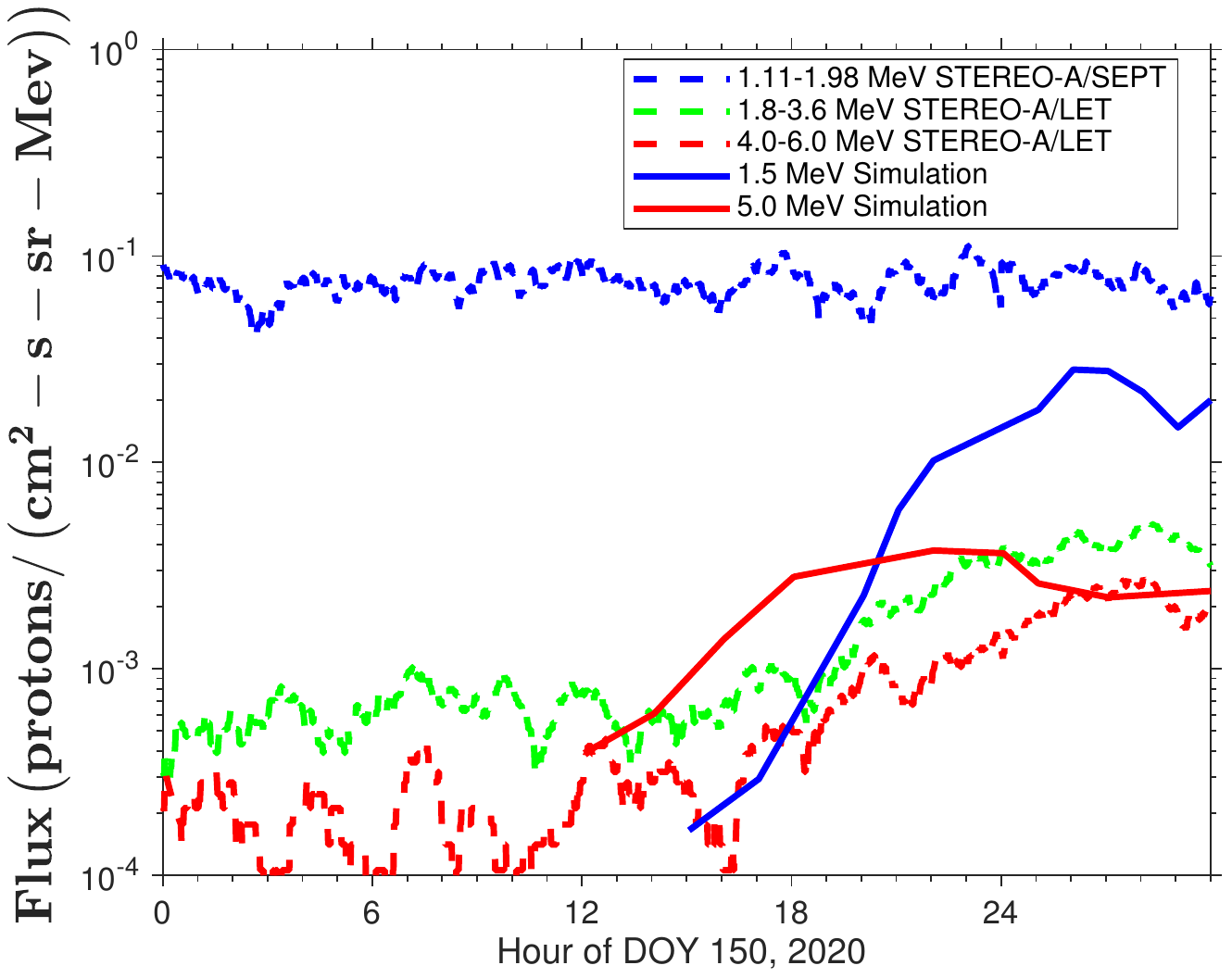}
	\caption{Time evolution of proton flux at 1.11\textendash1.98 MeV (blue dash curve), 1.8\textendash3.6 MeV (green dash curve) and 4.0\textendash6.0 MeV (red dash curve) from  \textit{STEREO-A} observations and proton flux  at 1.5 MeV (blue solid curve) and 8.0 MeV (red solid curve) from the simulation. Data of the observations are from the LET and SEPT instruments of \textit{STEREO-A}. 
		\label{fig:stereoA}}
\end{figure}

Our calculations for \textit{SOHO} do not show any SEP fluxes above the observation background level of the instrument on this spacecraft. The result is consistent with the observation of no SEP enhancement during the time period.

\section{Summary and Discussion}\label{sec:disc}

In this paper, a simulation has been carried out using time-backward stochastic differential equations to derive energetic particle intensities during SEP events from  the focus transport equation. A SEP source term is given in a form consistent with the theory of diffusive shock acceleration by a CME shock front. It only needs an input of the particle distribution function of accelerated SEPs at the shock front, which can be established once the local shock compression is known and the total number of injected particles is estimated. The code takes the input of coronal magnetic field configuration based on a data-driven PFSS model and propagates CME shock reconstructed from coronagraph observations. The simulation is applied to the 2020 May 29 SEP event observed by \textit{PSP}, while \textit{PSP} was not magnetically connected to the CME shock. We calculate the proton intensity-time profile at PSP and estimate the spectral index of ions between  $\sim \rm 2$ MeV and $16$ MeV and the path length of the field line experienced by particles at the event onset. Overall, our results are consistent with the observations. The flux of 2.2 MeV protons from the simulation and the \textit{PSP}  observation are in the same order of magnitude with the peak value of 0.32 (cm$^{2}$ s sr MeV)$^{-1}$
and 0.15 (cm$^{2}$ s sr MeV)$^{-1}$, respectively, whereas the simulated 2.2 MeV proton intensities increased 1.5 hours later than that from the \textit{PSP} observations.  The modeled flux of 5.0 MeV protons has a peak value  around $3.6\times 10^{-3}$ (cm$^{2}$ s sr MeV)$^{-1}$, which is between the  peak value of the 1.8\textendash3.6 MeV  and 4.0-6.0 MeV proton intensities measured by the LET on \textit{STEREO-A}. The estimated spectral index is 2.08 from the simulation, which is consistent with the slope of 2.18 from the \textit{PSP} observation. The estimated  equivalent path length of 0.70 au, which is slightly larger than 0.625 au from the \textit{PSP} observations estimated by \citet{chhiber2021magnetic}.  Even though the CME is slow, below the speed of the fully accelerated solar wind, it could drive a shock wave and accelerate particles in the low solar corona.  Because the heliocentric radial distance of \textit{PSP} was small and, in principle, a shock driven by a slow CME can only survive in the low corona before the CME sufficiently expands, the probability of a direct magnetic connection between a spacecraft and a mobile source of particles through the average magnetic field is quite low. The fact that SEPs were seen by \textit{PSP} requires the particles to propagate across magnetic field lines. A rate of about 10\% supergranular diffusion for the field line random walk is assumed in this paper to drive the particles' perpendicular diffusion \citep{zhang2017precipitation}. If the perpendicular transport of particles is through the random walk of magnetic field lines, we found that a fraction of supergranular diffusion at the base of the photosphere is sufficient to explain the particle diffusion that is needed for \textit{PSP} observations.

M.Z. acknowledges support from NASA grant 80NSSC19K1254.
L.A.B.  acknowledges support from NASA grant 80NSSC19K1235.
D.L. acknowledges support from NASA Living With a Star (LWS) programs NNH17ZDA001N-LWS and NNH19ZDA001N-LWS, the Goddard Space Flight Center Internal Scientist Funding Model (competitive work package) program and the Heliophysics Innovation Fund (HIF) program. This CME catalog is generated and maintained at the CDAW Data Center by NASA and The Catholic University of America in cooperation with the Naval Research Laboratory. SOHO is a project of international cooperation between ESA and NASA. Parker Solar Probe was designed, built, and is now operated by the Johns Hopkins Applied Physics Laboratory as part of NASA's Living with a Star (LWS) program. The data from the observation are downloaded from \url{http://spdf.gsfc.nasa.gov}. The computation facility used to run the simulations in the paper was supported by the National Science Foundation under Grant No. CNS 2016818.

%
%
%

\bibliography{ref}{}
\bibliographystyle{aasjournal}
\end{document}